%% file: main.tex
\documentclass[aps,prl,reprint,preprintnumbers,groupedaddress,amsmath,amssymb]{revtex4-2}

\usepackage{graphicx}
\usepackage{dcolumn}
\usepackage{bm}

\usepackage{physics}
\usepackage[range-phrase=--]{siunitx}
\usepackage{color}
\usepackage{xcolor}
\definecolor{fak4blue}{RGB}{0, 118, 175/89, 34, 2, 9}
\usepackage[colorlinks=true,citecolor=fak4blue,linkcolor=fak4blue,urlcolor=fak4blue]{hyperref}
\usepackage[noabbrev,nameinlink]{cleveref}
\usepackage{float}
\usepackage{multirow}
\usepackage{booktabs}
\usepackage{csquotes}
\usepackage{mathtools}
\usepackage{orcidlink}

\newcommand{\runinheading}[1]{\par\textit{#1---}\ignorespaces}

\crefname{appendix}{appendix}{appendices}
\Crefname{appendix}{Appendix}{Appendices}
\newcommand{\appendixheading}[2]{\par\refstepcounter{section}\label[appendix]{#2}\textit{Appendix~\thesection: #1---}\ignorespaces}

\hypersetup{
 pdftitle={Parallel Tempered Metadynamics for full QCD},
 pdfauthor={Eichhorn, Fuwa, Hoelbling, Varnhorst},
 bookmarksnumbered=true,
 bookmarksopen=true,
 bookmarksopenlevel=2
}

\begin{document}

\preprint{WUB/26-02}

\title{Parallel Tempered Metadynamics for full QCD}

\author{Timo Eichhorn \orcidlink{0000-0001-9370-3642}}
\email{timo.eichhorn@protonmail.com}
\author{Gianluca Fuwa \orcidlink{0000-0002-8195-4900}}
\email{gianluca.fuwa@uni-wuppertal.de}
\author{Christian Hoelbling \orcidlink{0000-0001-5715-1086}}
\email{hch@uni-wuppertal.de}
\author{Lukas Varnhorst \orcidlink{0000-0002-3718-0143}}
\email{varnhorst@uni-wuppertal.de}

\affiliation{Department of Physics, University of Wuppertal, Gaußstraße 20, 42119 Wuppertal, Germany}

\date{\today}

\begin{abstract}
We present an algorithm that addresses topological freezing in lattice QCD simulations by combining parallel tempering with collective-variable-based enhanced sampling methods, and apply it to a particularly challenging system with $N_f = 2$ staggered fermions. We find that the algorithm unfreezes the system, which is otherwise completely frozen for approximately $\num{40000}$ Molecular Dynamics Units with the Rational Hybrid Monte Carlo algorithm.
\end{abstract}

\maketitle

\input{include/sections/1}
\input{include/sections/2}
\input{include/sections/3}
\input{include/sections/4}
\input{include/sections/5}

\runinheading{Acknowledgments}
We thank Szabolcs Borsanyi and Stephan Dürr for helpful discussions. T.E. wishes to acknowledge numerous helpful discussions at the \enquote{Multi-canonical methods and lattice field theory} workshop in Trento.

T.E. and L.V. are supported by the Deutsche Forschungsgemeinschaft (DFG, grant No. HO 4177/1-1). Computations were carried out on the PLEIADES cluster at the University of Wuppertal, which was supported by the Deutsche Forschungsgemeinschaft (DFG, grant No. INST 218/78-1 FUGG) and the Bundesministerium für Forschung, Technologie und Raumfahrt (BMFTR).

\bibliography{literature}
\appendix

\onecolumngrid
\vspace{\topskip}
\begingroup
    \centering
    \large\textbf{End Matter}\par
\endgroup
\vspace{\topskip}
\twocolumngrid
\input{include/sections/6}

\end{document}

%% file: include/sections/1.tex
\runinheading{Introduction}
A major source of systematic uncertainty in contemporary lattice quantum chromodynamics (QCD) simulations stems from discretization artifacts \cite{FlavourLatticeAveragingGroupFLAG:2024oxs}. These cutoff effects can be substantially reduced by employing improved actions and observables, but a controlled continuum extrapolation involving multiple lattice spacings remains necessary for reliable error estimates.

At the same time, Markov chain Monte Carlo simulations become increasingly inefficient with finer lattice spacings. In particular, as the distinct topological sectors of the theory emerge towards the continuum, the integrated autocorrelation times of topological observables, and with them the number of updates required to produce decorrelated samples, have been observed to grow approximately as the fifth power of the inverse lattice spacing, or even faster \cite{DelDebbio:2004xh, Schaefer:2009xx, Schaefer:2010hu, Eichhorn:2023uge, Eichhorn:2025aph}. This issue, commonly referred to as topological freezing, has been known since the early days of lattice QCD \cite{Hoek:1987jg, Alles:1996vn}, but has become increasingly relevant as simulations have started targeting sub-percent precision. Consequently, the development of more efficient update algorithms has received renewed attention in recent years; see, e.g., \cite{Luscher:2011kk, Laio:2015era, Hasenbusch:2017unr, Albergo:2019eim, Kanwar:2020xzo, Cossu:2021bgn, Albandea:2021lvl, Bonanno:2024udh, Bonanno:2024zyn, Zhu:2025pmw, Bonanno:2025pdp}, as well as \cite{Finkenrath:2023sjg, Kanwar:2024ujc, Boyle:2024nlh, Finkenrath:2024ptc} and references therein.

In a previous paper \cite{Eichhorn:2023uge}, we proposed an algorithm, Parallel Tempered Metadynamics (PT-MetaD), and demonstrated that it can overcome topological freezing in two-dimensional $\mathrm{U(1)}$ and in four-dimensional $\mathrm{SU(3)}$ gauge theory. In this Letter, we apply PT-MetaD to full QCD and demonstrate that it samples topological sectors in a regime where a conventional (Rational) Hybrid Monte Carlo ((R)HMC) algorithm is fully frozen for approximately $\num{40000}$ Molecular Dynamics Units.

%% file: include/sections/2.tex
\runinheading{Method}
Lattice QCD simulations are based on numerically estimating a discretized path integral, which amounts to sampling $\mathrm{SU}(3)$-valued links $U$ according to the probability distribution
\begin{equation}
    p(U) = \frac{1}{Z} e^{-S(U)}, \qquad Z = \int \mathcal{D}[U]\, e^{-S(U)}.
\end{equation}
Here $\mathcal{D}[U]$ denotes the product of Haar measures over the link variables, and $S(U)$ is the action of the lattice theory. Conventionally, HMC-based algorithms \cite{Duane:1987de} are used for sampling, but their efficiency deteriorates for multimodal distributions, such as those encountered with the emergence of topological sectors in lattice QCD.

The method we propose combines parallel tempering \cite{PhysRevLett.57.2607} with collective-variable-based enhanced sampling, in particular Metadynamics \cite{Laio_2002}. It is applicable to systems that can be characterized by a probability distribution in one or more collective variables (CVs) with several high-probability regions separated by low probability regions.

In parallel tempering, we consider $n$ replicas of a system with identical state spaces, sampled with probability distributions $p_i$. Each replica is individually updated using a Markov chain that leaves the respective $p_i$ invariant. Additionally, swaps between adjacent replicas are proposed periodically, where a proposed swap of the configurations in replicas $i$ and $j$ is accepted with probability
\begin{equation}
    A_{i \leftrightarrow j}(U_i, U_j) = \min\biggl\{1, \frac{p_i(U_j) p_j(U_i)}{p_i(U_i) p_j(U_j)}\biggr\}.
\end{equation}
Here $U_k$ refers to the configuration in the $k^\mathrm{th}$ replica at the time the swap is proposed. As a result, the extended system is distributed according to the joint distribution
\begin{equation}
    p(U_1, \dots, U_n) \propto \prod_{i = 1}^n p_i(U_i).
\end{equation}
In the context of this Letter, we are primarily interested in the distribution $p_1$, with auxiliary distributions $\{p_i\}_{i \in \{2, \dots, n\}}$ to improve the sampling of $p_1$.

Typically, replicas operate at different couplings, and the probability distributions $p_i(U)$ can be defined by interpolating between the endpoints $p_1$ and $p_n$:
\begin{equation}
    p_i(U) \propto p_1(U) r(U)^{\alpha_i},
\end{equation}
where $0 = \alpha_1 < \alpha_2 < \dots < \alpha_n = 1$, and the interpolator $r(U)$ is defined as
\begin{equation}
    r(U) \propto \frac{p_n(U)}{p_1(U)}.
\end{equation}
We instead propose to define
\begin{equation}
    p_i(U) \propto p_1(U) p_{V}(\mathbf{s}(U))^{\alpha_i},
\end{equation}
where the interpolator $p_{V}(\mathbf{s}(U))$ depends on a set of collective variables $s_j(U)$. In standard parallel tempering, which interpolates between systems by varying a coupling that affects a number of fundamental degrees of freedom $N$, the number of replicas required for constant swap rates scales with $\sqrt{N}$ \cite{10.1063/1.1507776,Atchade2011}. In PT-MetaD, we instead introduce $N$ collective variables that are specifically designed to facilitate transitions between different high-probability regions in a system. While this invalidates the basic assumptions leading to the $\sqrt{N}$ scaling, we expect it to hold approximately unless the bias potential is extremely peaked. Based on the knowledge of $p_1(U)$ and $p_{V}(\mathbf{s}(U))$, one can compute the expected swap rate (see \Cref{app:swap_rate_model}), and of course one can check the swap rate in the actual simulation.

Note that the interpolator can be expressed in terms of a bias potential $V(\mathbf{s})$:
\begin{equation}
    p_{V}(\mathbf{s}) \propto e^{-V(\mathbf{s})}.
    \label{eq:bias_interpolator}
\end{equation}
Thus, the bias potential required to realize a prescribed target distribution $p_{\mathrm{tg}}(\mathbf{s})$ is, up to an irrelevant additive constant, given by
\begin{equation}
    V(\mathbf{s}) = \ln\biggl(\frac{p(\mathbf{s})}{p_{\mathrm{tg}}(\mathbf{s})}\biggr),
\end{equation}
where $p(\mathbf{s})$ is the marginal distribution over the CVs with respect to the unbiased distribution. Common target distributions are the uniform distribution or well-tempered target distributions, where $p_{\mathrm{tg}}(\mathbf{s}) \propto p(\mathbf{s})^{1/\gamma}$ (with $\gamma \geq 1$). While $p(\mathbf{s})$ is generally not known a priori, Metadynamics and related enhanced sampling methods provide a way to iteratively reconstruct the marginal distribution, or equivalently, the free energy surface $\mathrm{FES}(\mathbf{s})$ of the collective variables $\mathbf{s}$ \cite{Laio_2002, PhysRevLett.100.020603, PhysRevLett.113.090601, Invernizzi_2020}. Although we use an unbiased $p_1$ in this Letter, it is in principle possible to bias it too for reducing the variance of specific observables. For more detailed discussions in the context of lattice gauge theory, we refer to previous publications \cite{Laio:2015era, Bonati:2017nhe, Eichhorn:2023uge, Eichhorn:2026ehi}.

%% file: include/sections/3.tex
\runinheading{Simulation setup}
We simulate two mass-degenerate flavors of staggered fermions \cite{Kogut:1974ag, Banks:1976ia, Susskind:1976jm} with $V/a^4 = 24^4$, $am = 0.02$, four stout smearing \cite{Morningstar:2003gk} steps with smearing parameter $\rho = 0.125$, and the doubly blocked Wilson (DBW2) action \cite{Takaishi:1996xj}. The DBW2 action is a popular choice for applications where a clear separation of the topological sectors is desired already at relatively coarse lattice spacings \cite{Orginos:2001xa, DeGrand:2002vu, Noaki:2002ai, Aoki:2005ga, Borsanyi:2020mff, Borsanyi:2023tdp, Borsanyi:2023wno, Borsanyi:2024xrx, Borsanyi:2025kiv}. This very same feature, however, causes simulations with the DBW2 action to experience topological freezing at relatively coarse lattice spacings.

For each ensemble, we perform both conventional RHMC \cite{Kennedy:1998cu, Clark:2003na} and PT-MetaD simulations with the parameters given in \Cref{tab:simulation_parameters}. The RHMC trajectories have a length of four Molecular Dynamics Units (MDUs), and are integrated with a fourth-order minimum norm integrator \cite{OMELYAN2003272} and a step size of \num{0.2} for all forces.
\begin{table}[h]
    \caption{Overview of our simulation parameters. All simulations use $N_f = 2$ staggered fermions with four steps of stout smearing. Scale setting was performed using $t_0$ \cite{Luscher:2010iy, Bruno:2013gha}.}
    \label{tab:simulation_parameters}
    \begin{ruledtabular}
        \begin{tabular}{ccccc}
            $L/a$               & $am$                  & $\beta$               & $a$ [\si{\femto\m}]                      & Update algorithm \\ \midrule
            \multirow{4}{*}{24} & \multirow{4}{*}{0.02} & \multirow{2}{*}{0.95} & \multirow{2}{*}{\num{0.0633 \pm 0.0018}} & RHMC \\
                                &                       &                       &                                          & PT-MetaD \\
                                &                       & \multirow{2}{*}{1.15} & \multirow{2}{*}{\num{0.0369 \pm 0.0067}} & RHMC \\
                                &                       &                       &                                          & PT-MetaD
        \end{tabular}
    \end{ruledtabular}
\end{table}
For the PT-MetaD simulations, we use two replicas and a single collective variable $Q_{\mathrm{CV}}$, defined as the clover-based topological charge after six stout smearing steps with smearing parameter $\rho = 0.125$. Individually, each replica is updated with the RHMC algorithm using the same parameters as for the conventional reference runs. The first replica is sampled according to the unmodified physical distribution, whereas the action of the second replica includes a topological bias potential $V(Q_{\mathrm{CV}})$. Accordingly, the swap probabilities are given by
\begin{equation}
    A_{1 \leftrightarrow 2}(U_1, U_2) = \min\Biggl\{1, \frac{\exp\bigl[V\bigl(Q_\mathrm{CV}(U_2)\bigr)\bigr]}{\exp\bigl[V\bigl(Q_\mathrm{CV}(U_1)\bigr)\bigr]}\Biggr\}.
\end{equation}
Importantly, this expression does not involve the physical action or any fermionic contributions, so the evaluation of the action difference during the accept-reject step adds negligible computational overhead to the overall simulation. Furthermore, since the fermions and the CV use the same smearing parameter, and all forces are integrated with the same integrator on the same timescale, it is possible to further reduce the computational overhead by merging the stout force recursions for the fermionic and the bias force terms.

To obtain an initial estimate of the free energy surfaces, we perform well-tempered Metadynamics simulations \cite{PhysRevLett.100.020603}. Afterwards, the bias potentials are obtained by using singular spectrum analysis \cite{VAUTARD1989395} to extract the periodic part of the negative free energy surface, which only contains the barriers between topological sectors. This preserves the relative weights of the sectors in the biased distribution, thus increasing the overlap between $p_1$ and $p_2$, and thereby the swap rates and efficiency of the algorithm, as shown in \cite{Eichhorn:2023uge}. Further details on how the bias potentials were obtained are given in \Cref{app:potential_buildup}.

%% file: include/sections/4.tex
\begin{figure}[h]
    \centering
    \includegraphics{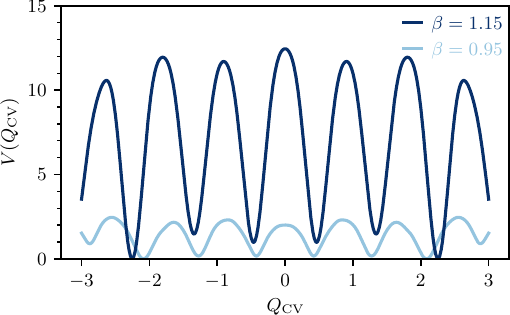}
    \caption{Bias potentials used in the PT-MetaD runs. The potentials were obtained from well-tempered Metadynamics buildup runs and subsequently filtered by singular spectrum analysis to only retain the oscillatory component associated with intersector barriers (see \Cref{app:potential_buildup} for details).}
    \label{fig:bias_potential_ssa_comparison}
\end{figure}

\runinheading{Results}
In \Cref{fig:bias_potential_ssa_comparison} we show the bias potentials used for our two runs. The buildup simulations ran for \num{12000} and \num{16000} trajectories for the coarse and fine lattice, respectively. Clearly, the intersector barriers are very different for the two couplings we investigate. The residual fluctuations indicate that the potentials are not fully equilibrated, which can reduce tunneling and swap acceptance rates, but does not introduce any bias on expectation values in the physical replica.
To check how well the two algorithms perform, we monitor the plaquette $P$ and the clover-based topological charge $Q$ after $30$ steps of stout smearing with smearing parameter $\rho = 0.125$. Additionally, we compute the topological susceptibility based on rounding $Q$ to the nearest integer.

\begin{figure}[h]
    \centering
    \includegraphics{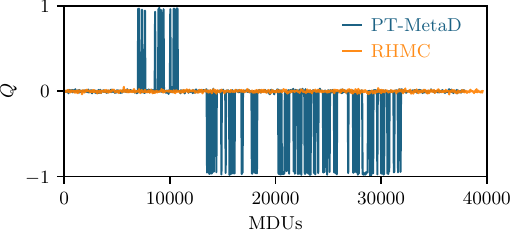}
    \caption{Time series of the topological charge at $\beta = 1.15$ using the RHMC and the PT-MetaD algorithm. The RHMC algorithm remains within the zero sector, whereas PT-MetaD repeatedly samples multiple sectors.}
    \label{fig:qclov_compare_dbw2_1.15}
\end{figure}

For our purposes, the interesting case is the $\beta = 1.15$ system. The barrier heights of $\mathcal{O}(10)$ between topological sectors indicate that conventional algorithms will have an extremely small transition probability between topological sectors. Indeed, \Cref{fig:qclov_compare_dbw2_1.15} shows that the topological charge is completely frozen for $9900$ RHMC trajectories of total length $39600$ MDUs, whereas PT-MetaD repeatedly tunnels between sectors.

\begin{figure}[h]
    \centering
    \includegraphics{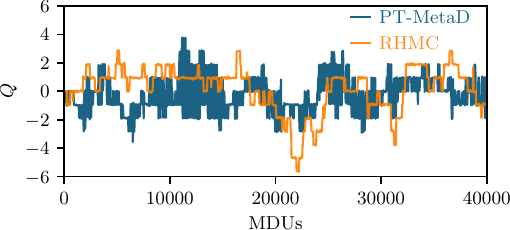}
    \caption{Time series of the topological charge at $\beta = 0.95$ using the RHMC and the PT-MetaD algorithm. Both algorithms sample multiple sectors, but PT-MetaD exhibits more frequent transitions and a shorter integrated autocorrelation time of $Q^2$ (see \Cref{tab:integrated_autocorrelation_times}).}
    \label{fig:qclov_compare_dbw2_0.95}
\end{figure}

As a crosscheck, we also show the time series for the system at $\beta=0.95$ in \Cref{fig:qclov_compare_dbw2_0.95}. As the small barrier height of the bias potential in \Cref{fig:bias_potential_ssa_comparison} suggests, the RHMC tunnels frequently, albeit not as frequently as PT-MetaD.
\Cref{tab:expectation_values} and \Cref{tab:integrated_autocorrelation_times} contain the results and integrated autocorrelation times of our observables for the four runs, as determined with the $\Gamma$-method \cite{Wolff:2003sm,Ramos:2020scv}. For the coarse lattice, both observables are in agreement, although the PT-MetaD estimate for the topological susceptibility and the smeared plaquette have errors that are smaller by factors $\sim \num{2.62}$ and $\sim \num{1.06}$, respectively. Since the RHMC is completely frozen on the fine lattice, we cannot provide a reasonable estimate of the susceptibility. The topological freezing also visibly affects the smeared plaquette, which shows an almost $10 \sigma$ deviation from the PT-MetaD estimate.
\begin{table}[h]
    \caption{Expectation values of the smeared plaquette $P$ and the dimensionless topological susceptibility $\chi_\mathrm{top} V$. At each $\beta$, the longer time series was truncated such that the RHMC and PT-MetaD results are based on the same number of measurements $n_\mathrm{trunc}$, separated by \SI{40}{MDUs}.}
    \label{tab:expectation_values}
    \begin{ruledtabular}
        \begin{tabular}{ccccc}
            $\beta$              & Algorithm & $n_\mathrm{trunc}$ & $P$                             & $\chi_\mathrm{top} V$ \\ \midrule
           \multirow{2}{*}{0.95} & RHMC      & \num{1063}         & \num{0.99950150 \pm 0.00000084} & \num{2.73 \pm 0.93}   \\
                                 & PT-MetaD  & \num{1063}         & \num{0.99950140 \pm 0.00000079} & \num{2.17 \pm 0.36}   \\
           \multirow{2}{*}{1.15} & RHMC      & \num{947}          & \num{0.99972054 \pm 0.00000127} & 0                     \\
                                 & PT-MetaD  & \num{947}          & \num{0.99973404 \pm 0.00000052} & \num{0.127 \pm 0.033}
        \end{tabular}
    \end{ruledtabular}
\end{table}
\begin{table}[h]
    \caption{Integrated autocorrelation times of different observables, expressed in MDUs and estimated from the full available time series of $n_\mathrm{meas}$ measurements, separated by \SI{40}{MDUs}. Here, $Q_1$ and $Q_2$ denote the topological charges measured in the physical and biased replicas, respectively.}
    \label{tab:integrated_autocorrelation_times}
    \begin{ruledtabular}
    \begin{tabular}{ccccc}
            $\beta$              & Algorithm & $n_\mathrm{meas}$ & $\tau_\mathrm{int}(\chi_\mathrm{top})$ & $\tau_\mathrm{int}((Q_1+Q_2)^2)$ \\ \midrule
           \multirow{2}{*}{0.95} & RHMC      & \num{1063}        & \num{764 \pm 488}                      & N/A                              \\
                                 & PT-MetaD  & \num{1100}        & \num{392 \pm 208}                      & \num{481 \pm 269}                \\
           \multirow{2}{*}{1.15} & RHMC      & \num{990}         & --                                     & N/A                              \\
                                 & PT-MetaD  & \num{947}         & \num{180 \pm 76}                       & \num{1591 \pm 1037}
    \end{tabular}
    \end{ruledtabular}
\end{table}

Regarding the computational costs, our PT-MetaD runs used a factor $\sim \num{2.3}$ more CPU time compared to the RHMC runs. Taking the computational overhead of the additional replica into account, this corresponds to an efficiency gain of $\sim \num{3.0}$ in CPU time for the topological susceptibility on the coarse lattice. Even when including the one-time cost of building up the bias potential, PT-MetaD is still more efficient by a factor $\sim \num{1.82}$. In any case, it is clear that PT-MetaD is at the very least comparable in CPU efficiency, even for a system where topology changes occur rather frequently. Furthermore, we would like to emphasize that both the bias potential buildup and the actual simulation are trivially parallelizable in the number of walkers (see \Cref{app:potential_buildup}) and replicas, respectively. We have also not included measurements on the tunneling replica for any of our observables, which would further increase their precision.

%% file: include/sections/5.tex
\runinheading{Conclusion}
In this Letter, we have demonstrated that Parallel Tempered Metadynamics is able to unfreeze simulations in full lattice QCD for a system which is completely frozen with a conventional Hybrid Monte Carlo-based algorithm for $\sim 4 \times 10^4$ Molecular Dynamics Units. Even for a system where the conventional algorithm is not frozen and tunneling events are relatively frequent, the reduced autocorrelation times of the topological susceptibility more than compensate any computational overhead.

Whether these observations imply a better continuum scaling, as previously observed in the case of 2-dimensional $\mathrm{U(1)}$ gauge theory \cite{Eichhorn:2023uge}, and whether this can be achieved without adjusting (the smearing parameter of) the collective variable, remains to be seen.

We are currently in the progress of applying the algorithm to the computation of the topological susceptibility at high temperatures and small lattice spacings, which exhibits both topological freezing and a suppression of non-trivial topological sectors. More broadly, it would be interesting to see whether the underlying strategy may generically be useful for multimodal systems where appropriate collective variables can be found.

%% file: include/sections/6.tex
\appendixheading{Potential buildup}{app:potential_buildup}
We perform well-tempered Metadynamics \cite{PhysRevLett.100.020603} simulations with $\gamma=1.1$ to obtain initial estimates of the free energy surfaces. The simulations use four parallel walkers \cite{doi:10.1021/jp054359r}, with \num{3000} trajectories per walker for the $\beta = \num{0.95}$ run and \num{4000} trajectories per walker for the $\beta = \num{1.15}$ run. The estimates are stored in histograms with bin widths of $0.01$. Additionally, we include measurements during trajectories in a way that is based on the Recycled HMC \cite{Nishimura_2020}, similar to the process described in \cite{Eichhorn:2026ehi}. However, we only perform a single accept-reject step at the end of the trajectory, discarding all measurements during the trajectory upon rejection. This introduces a bias in the estimate of the free energy surface, which is negligible compared to statistical fluctuations at the given level of precision and high acceptance rates. As already pointed out in the main text, an inexact estimate of the free energy surface does not affect the correctness of our algorithm, only its efficiency. Otherwise, each simulation utilizes the same setup as the RHMC simulations, as described in the main text.

\begin{figure}[h]
    \centering
    \includegraphics{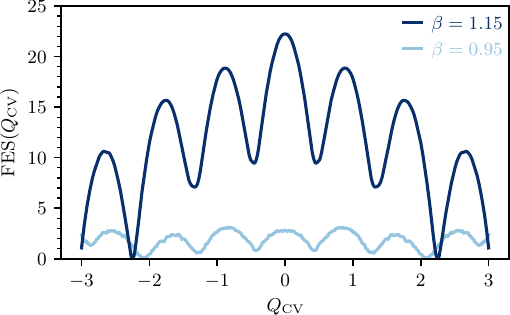}
    \caption{The estimated free energy surfaces obtained via \Cref{eq:fes_estimate}.}
    \label{fig:fes_comparison}
\end{figure}

During the buildup phase, we track the Metadynamics bias potential $V_\mathrm{meta}(\mathbf{s}, t)$ at simulation time $t$, from which we then construct the estimated free energy surface at the final time $T$ \cite{doi:10.1021/acs.jctc.9b00867}:
\begin{equation}
    -\mathrm{FES}(\mathbf{s}) = \ln \Biggl(\frac{\sum_{t'}^T g\bigl(\mathbf{s} - \mathbf{s}(t')\bigr) w\bigl(\mathbf{s}(t'), t'\bigr)}{\sum_{t'}^T w\bigl(\mathbf{s}(t'), t'\bigr)}\Biggr).
    \label{eq:fes_estimate}
\end{equation}
Here $g\bigl(\mathbf{s}\bigr)$ denotes a normalized Gaussian of variance 0.02, and the reweighting factors $w\bigl(\mathbf{s}(t), t\bigr)$ are calculated based on the Tiwary scheme \cite{doi:10.1021/acs.jctc.9b00867}:
\begin{equation}
    w\bigl(\mathbf{s}(t'), t'\bigr) = \frac{e^{V_\mathrm{meta}(\mathbf{s}(t'), t')}}{\langle e^{V_\mathrm{meta}(\mathbf{s}, t')}\rangle_\mathbf{s}}
\end{equation}
where $\langle \cdot \rangle_s$ denotes the uniform average over the explored range of CVs. The resulting estimates of the free energy surfaces are displayed in \Cref{fig:fes_comparison}.

We obtain the final bias potentials for our production runs by applying singular spectrum analysis \cite{VAUTARD1989395} to extract the periodic parts of the negative free energy surfaces. The resulting potentials only contain the barriers between topological sectors. This preserves the relative weights of the sectors in the biased distribution, thus increasing the overlap between $p_1$ and $p_2$, and with it the swap rates and efficiency of the algorithm, as shown in \cite{Eichhorn:2023uge}.

\appendixheading{A theoretical model for swap rates}{app:swap_rate_model}
In a general parallel tempering setup, the expected swap rate between two replicas can be computed according to
\begin{equation}
    \begin{aligned}
        \langle A_{i \leftrightarrow j} \rangle = \int \int & \mathrm{d}U_i\, \mathrm{d}U_j\, p_i(U_i) p_j(U_j)
        \\ \times& \min\biggl\{1, \frac{p_i(U_j) p_j(U_i)}{p_i(U_i) p_j(U_j)} \biggr\}.
    \end{aligned}
\end{equation}
For PT-MetaD with two replicas, this expression may be written as
\begin{equation}
    \begin{aligned}
        \langle A_{1 \leftrightarrow 2} \rangle = \int_R \int_R & \mathrm{d}Q_1 \mathrm{d}Q_2\, p(Q_1) p(Q_2) p_V(Q_2)
        \\ \times& \min\biggl\{1, \frac{p_V(Q_1)}{p_V(Q_2)}\biggr\},
    \end{aligned}
\end{equation}
where $R$ is the range over which we have determined our bias potential.
Neither $p(Q)$ nor $p_V(Q)$ are known a priori, but the extraction of the free energy landscape in the bias potential buildup allows us to estimate both. In fact, $p(Q)$ is the marginal distribution with respect to $Q$ of the unbiased replica, for which we have a direct estimate, up to a normalization constant:
\begin{equation}
    p(Q)\propto e^{-\mathrm{FES}(Q)}.
\end{equation}
The ratio of marginal distributions $p_V(Q)$ is given, up to another normalization constant, in terms of the bias potential via \Cref{eq:bias_interpolator}. The normalization constants are fixed by 
demanding unit overall probability, i.e.,
\begin{equation}
    \int_R \mathrm{d}Q\, p(Q) = \int_R \mathrm{d}Q\, p(Q) p_V(Q)=1.
\end{equation}
The expected swap rates of  our runs are thus \SI{64.3}{\percent} for $\beta = 0.95$ and  \SI{31.0}{\percent} for $\beta = 1.15$, which is in reasonable agreement with the actually observed swap rates of \SI{68.6}{\percent} and \SI{28.6}{\percent}, respectively.

Note that while the ratio $p_V(Q)$ is exactly the one we used in actual simulations, we only have an estimate for the marginal distribution $p(Q)$. Thus the difference between expected and measured swap rates is caused, apart from obvious statistical fluctuations, by inaccuracies in the estimation of the free energy surface. The agreement between the expected and observed swap rates can thus be used as an indicator of how accurately the free energy surface was estimated.

\begin{figure}[h]
    \centering
    \includegraphics{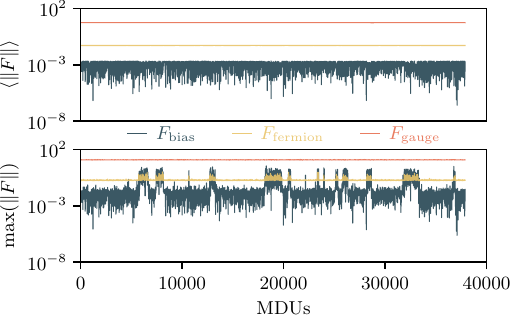}
    \caption{Time series of the average and maximum RHMC forces for the biased replica of the PT-MetaD run at $\beta = \num{1.15}$.}
    \label{fig:rhmc_forces}
\end{figure}

\appendixheading{RHMC forces}{app:rhmc_forces}
\Cref{fig:rhmc_forces} shows the average and maximum forces during the RHMC evolution on the biased replica of the PT-MetaD simulation at $\beta = \num{1.15}$. It is interesting to note that while the mean and maximum forces are compatible with each other within one order of magnitude for both the gauge and fermionic force terms, the discrepancy is far larger for the bias force. This suggests that the force incurred by $Q_\mathrm{CV}$ is localized. These local spikes in the bias force occur predominantly when the system traverses the intersector regions, and are also visible in the fermionic force, although to a lesser extent.
